\documentclass[aip,jcp,reprint,amsmath,amssymb,floatfix,citeautoscript]{revtex4-2}
\usepackage{cancel}
\usepackage{amsmath}
\usepackage{physics}
\usepackage{footmisc}
\usepackage{graphicx}
\usepackage{ulem}
\usepackage{amssymb}
\usepackage{sidecap}
\usepackage{amsthm}
\usepackage{bm}
\usepackage{dcolumn}
\usepackage{braket}
\usepackage{longtable}
\usepackage{multirow}
\usepackage{booktabs}
\usepackage{footnote}
\usepackage{ragged2e}
\usepackage{txfonts}

\usepackage{color}
\usepackage[usenames,dvipsnames]{xcolor}
\definecolor{myblue}{rgb}{0,0,1}
\usepackage[breaklinks=true,colorlinks=true,linkcolor=myblue,urlcolor=myblue,citecol  or=myblue]{hyperref}

\newcommand{\wavenum}{cm$^{-1}$}
\newcommand{\eps}{{\varepsilon}}

\newcommand{\vari}{{\mathcal{V}}}
\newcommand{\vm}{{\bm{m}}}
\newcommand{\vn}{{\bm{n}}}
\newcommand{\statem}{\ket{\vm}}
\newcommand{\staten}{\ket{\vn}}

\begin{document}
\title{Vibrational heat-bath configuration interaction}
\author{Jonathan H. Fetherolf}
\affiliation{Department of Chemistry, 
Columbia University, New York, New York 10027, USA}
\author{Timothy C. Berkelbach}
\email{tim.berkelbach@gmail.com}
\affiliation{Center for Computational Quantum Physics, Flatiron Institute, New York, New York 10010, USA}
\affiliation{Department of Chemistry, 
Columbia University, New York, New York 10027, USA}

\begin{abstract}
We introduce vibrational heat-bath configuration interaction (VHCI) as an
accurate and efficient method for calculating vibrational eigenstates of
anharmonic systems.  Inspired by its origin in electronic structure theory, VHCI
is a selected CI approach that uses a simple criterion to identify important
basis states with a pre-sorted list of anharmonic force
constants.  Screened second-order perturbation theory and simple extrapolation
techniques provide significant improvements to variational energy estimates.  We
benchmark VHCI on four molecules with 12 to 48 degrees of freedom and use
anharmonic potential energy surfaces truncated at fourth and sixth order.  For
all molecules studied, VHCI produces vibrational spectra of tens or hundreds of
states with sub-wavenumber accuracy at low computational cost.
\end{abstract}

\maketitle

\section{Introduction}
Precise computational predictions of the vibrational structure of molecules and
solids requires the inclusion of anharmonic effects, which correspond to
interactions between harmonic normal modes.  When treated quantum mechanically,
this requires the accurate solution of the vibrational Schr\"odinger equation on
a high-dimensional potential energy surface.  Like in electronic structure
theory, a hierarchy of wavefunction-based methods are commonly employed to avoid
the exponential cost associated with an exact quantum solution; such methods
include vibrational self-consistent field
theory~\cite{B.Gerber2002,Chaban1999,Carter1997,Roy2013} vibrational
perturbation theory\cite{Barone2005,Christiansen2003,Sibert1987}, vibrational
coupled-cluster theory~\cite{Banik2008,Seidler2009,Christiansen2004}, and
vibrational configuration interaction
(VCI)\cite{Bowman1979,Thompson1980,Christoffel1982}.  These methods rely on the
accuracy of the harmonic approximation and alternative methods have recently
been developed to achieve quantitative accuracy for strongly anharmonic
vibrations.  Such methods include the nonproduct quadrature
approach\cite{Avila2011}, reduced rank block power method (RRBPM), which uses a
tensor factorization of the vibrational wavefunction~\cite{Leclerc2014}, and
adaptive VCI (A-VCI), which accelerates the VCI process using nested basis
functions~\cite{Odunlami2017}.

For strongly correlated electronic systems, the past two decades have seen the
development of powerful computational methods that could be brought to bear on
strongly anharmonic vibrational systems.  For example, a vibrational density
matrix renormalization group approach was recently developed and applied to
systems with up to sixteen atoms~\cite{Baiardi2017}. In this work, we are
primarily interested in selected configuration
interaction~\cite{Huron1973,Buenker1974,Harrison1991}, which began in the 1970s
with CI by perturbatively selecting iteratively~(CIPSI)~\cite{Huron1973} and has
experienced renewed interest in the form of adaptive configuration
interaction~\cite{Schriber2016}, adaptive sampling CI
(ASCI)~\cite{Tubman2016,Tubman2020}, and heat-bath CI
(HCI)~\cite{Holmes2016,Holmes2017,Sharma2017,Li2018}, among others.  In CIPSI,
many-body basis states are added to the variational CI space based on a
first-order approximation of their wavefunction coefficients.  Unlike CIPSI,
which considers all states belonging to the first-order interacting space, ASCI
only considers connections from the variational basis states that have
sufficiently large wavefunction amplitudes.  HCI uses a different selection
criterion that allows it to exploit the fact that many Hamiltonian matrix
elements are identical in magnitude; a presorting of these matrix elements (the
two-electron repulsion integrals) enables fast and efficient identification of
new basis states.  In almost all selected CI calculations, once the variational
space is suitably converged, a second-order perturbation theory (PT2) correction
is added to the variational energy.  Numerous recent studies have demonstrated
their power: HCI was selected as the most accurate method among twenty for a
study of transition metal atoms and their oxides~\cite{Williams2020}, HCI was
used to provide almost exact energies of the Gaussian-2 dataset~\cite{Yao2020},
and both HCI and ASCI have been used in a recent comparative study of the
ground-state energy of benzene~\cite{Eriksen2020}.

A variety of selected CI approaches have been developed for vibrational
problems~\cite{Rauhut2007,Neff2009,Carbonniere2010,Sibaev2016}, including
a vibrational CIPSI~\cite{Scribano2008} and a recent vibrational ASCI (VASCI)~\cite{Lesko2019}.
Motivated by HCI's strong performance and computational efficiency, in this work
we present a vibrational HCI (VHCI).  The differences between electronic and vibrational
problems, especially their Hilbert spaces and Hamiltonian forms, necessitate a number
of new developments in order for VHCI to enjoy the same advantages as electronic HCI; we describe 
these developments in Sec.~\ref{sec:theory}.  In Sec.~\ref{sec:results}, we present VHCI plus
perturbation theory results for molecules containing between 12 and 48 degrees of freedom,
calculating tens to hundreds of excited state energies.
By comparing to other literature results, we demonstrate that VHCI is a highly
accurate and efficient approach for large molecular systems with strong anharmonicity,
typically achieving sub-wavenumber accuracy with modest computational resources.

\section{Theory}
\label{sec:theory}

\subsection{Vibrational heat-bath configuration interaction}
\label{ssec:VHCI}
Expressed in terms of mass-weighted normal mode coordinates $Q_i$ with frequencies $\omega_i$,
the nuclear Hamiltonian is given by
\begin{equation}
    H =
    \frac{1}{2}\sum_{i=1}^D \left[ -\frac{\partial^2}{\partial Q_i^2} + \omega_i^2 Q_i^2 \right]
    + V_\mathrm{an}(Q_1,Q_2,\ldots,Q_D),
\end{equation}
where $D = 3N-6$ is the number of normal mode degrees of freedom and $V_\mathrm{an}$ is the anharmonic
part of the electronic ground state
potential energy surface (PES).  
We assume that the anharmonic part of the PES has a normal mode expansion
\begin{equation}
\begin{split}
    V_\mathrm{an}(\bm{Q}) &= \sum_{i\leq j \leq k} V_{ijk}Q_iQ_jQ_k 
        + \sum_{i\leq j \leq k \leq l} V_{ijkl}Q_iQ_jQ_kQ_l + \ldots \\
    &= \sum_{i\leq j \leq k} W_{ijk}\bar{Q}_i\bar{Q}_j\bar{Q}_k
        + \sum_{i\leq j \leq k \leq l} W_{ijkl}\bar{Q}_i\bar{Q}_j\bar{Q}_k\bar{Q}_l + \ldots \\
    \end{split}
    \label{eq:anharm}
\end{equation}
where the anharmonic force constants are partial derivatives
of the PES, e.g.~$V_{ijk}=(\partial^3V_\mathrm{an}/\partial Q_i \partial Q_j \partial Q_k)_{\bm{Q}=0}$.
In the second line of Eq.~(\ref{eq:anharm}), we define $\bar{Q}_i = (a_i^\dagger + a_i)$
and
\begin{equation}
W_{ijk\ldots} = \frac{V_{ijk\ldots}}{\sqrt{2^p \omega_i \omega_j \omega_k \ldots}},
\end{equation}
where $p$ is the order of the anharmonicity.
In practice the anharmonic PES expansion is truncated, commonly at fourth or sixth order.
In this work, we neglect the Coriolis rotational coupling, but such a term can be straightforwardly 
included.

In vibrational HCI, we choose to work in the basis of Hartree product states 
$|\bm{n}\rangle = |n_1,n_2,\ldots,n_D\rangle$ formed from the
harmonic oscillator orbitals $\phi_{n_i}(Q_i) = \langle Q_i|n_i\rangle$,
leading to the wavefunction
\begin{equation}
    \ket{\Psi} = \sum_{\vn\in \vari} c_\vn \staten.
    \label{eq:VCI}
\end{equation}
In VCI, the variational space $\vari$ is 
commonly truncated by excitation level, for example by limiting the number of 
vibrational quanta per product state or per mode.  However, this approach is not always efficient and
may lead to a large Hilbert space with many product states that contribute
minimally to the VCI solution.  This issue of inefficient addition of states is
addressed by selected VCI methods, which select states for inclusion in $\vari$ by
criteria other than excitation level as discussed in the Introduction, and here we focus on the HCI criterion.

We briefly recall that in the variational stage of HCI, 
basis state $\statem$ is added to $\vari$ if $|H_{\vm \vn} c_\vn| \ge \varepsilon_1$, where
$\varepsilon_1$ is a user-defined convergence parameter.  In electronic structure,
the matrix elements $H_{\vm\vn}$ arising from double excitations depend only on the
identity of the four orbitals involved.  Therefore, many of the Hamiltonian
matrix elements are identical in magnitude.  This is not true
in vibrational structure because the modes can have occupation numbers greater than one.
For example, consider an anharmonic PES with cubic and quartic terms.  The Hamiltonian matrix elements between
product states $\statem$ and $\staten$ that differ in their occupancy by one quantum
in mode $i$, by two quanta in modes $i$ and $j$ (including $i=j$), and so on,
are given by
\begin{subequations}
\begin{align}
\label{eq:single}
H_{\vm\vn}^{(i)} &= \sum_j W_{ijj} \langle \vm| \bar{Q}_i \bar{Q}_j^2 |\vn\rangle \\ 
\label{eq:double}
H_{\vm\vn}^{(i,j)} &= \sum_{k} W_{ijkk} \langle \vm | \bar{Q}_i \bar{Q}_j \bar{Q}_k^2 |\vn \rangle \\
H_{\vm\vn}^{(i,j,k)} &= W_{ijk} \langle \vm | \bar{Q}_i \bar{Q}_j \bar{Q}_k | \vn \rangle \\
H_{\vm\vn}^{(i,j,k,l)} &= W_{ijkl} \langle \vm | \bar{Q}_i \bar{Q}_j \bar{Q}_k \bar{Q}_l | \vn \rangle.
\end{align}
\end{subequations}
The factors $\langle \vm | \bar{Q}_i \ldots |\vn\rangle$ that are not zero are
products of terms of the form $\sqrt{n_i}$ and these factors are responsible for
the differentiation of most matrix elements in the Hamiltonian, precluding an
efficient evaluation of the usual HCI criterion.  
For example, if the product states $\statem$ and $\staten$ differ in the
occupancy of three modes $(i,j,k)$, then 
\begin{subequations}
\begin{align}
H_{\vm\vn}^{(i+,j+,k+)} &= W_{ijk} \sqrt{m_i m_j m_k} \\
H_{\vm\vn}^{(i+,j+,k-)} &= W_{ijk} \sqrt{m_i m_j n_k} \\
H_{\vm\vn}^{(i+,j-,k-)} &= W_{ijk} \sqrt{m_i n_j n_k}
\end{align}
\end{subequations}
and so on, where the notation indicates that the occupation of a given mode in $\statem$ is
bigger or smaller than in $\staten$.  Therefore, in addition to the \textit{identity} of the modes
with different occupations, the VCI matrix elements depend on (1) whether the occupations are
bigger/smaller on the bra/ket side, and (2) the overall excitation level.
However, because these latter factors $\langle \vm | \bar{Q}_i \ldots |\vn\rangle$
are products of terms of the form $\sqrt{n_i}$, they are of order unity 
(at least when the excitation level is reasonably low) and can be reasonably neglected for the purpose
of estimating the magnitude of the Hamiltonian matrix element.  Doing so leads to a significant
reduction in the number of unique numbers to be considered.

Specifically, let us define $W_i = \sum_j (2+\delta_{ij})W_{ijj}$ and 
$W_{ij} = \sum_{k} (2+\delta_{jk}+\delta_{ij}\delta_{jk}) W_{ijkk}$, which approximately account for
single and double mode excitations.
Then, given an approximate wavefunction of the form (\ref{eq:VCI}) expanded
over some space $\vari$, we propose to add state $\statem$ to the variational
space if $\statem$ and $\staten$ differ in their occupancy by one quantum
in mode $i$, by two quanta in modes $i$ and $j$ (including $i=j$), and so on,
and
\begin{equation}
\label{eq:VHCI_crit}
\left| W_{ijk\ldots} c_\vn \right| \ge \varepsilon_1.
\end{equation}
The above criterion can be implemented 
very efficiently by pre-sorting the matrix elements $W_{ijk\ldots}$, of which
there are at most a polynomial number, e.g.~$O(D^4)$ for a quartic PES.

The ground-state VHCI algorithm is performed as follows:
\begin{enumerate}
    \item Sort the list of $W_{ijk\ldots}$, largest to smallest.
    \item Initialize the space $\vari$ according to a small total number of quanta.
    \item Build the Hamiltonian matrix in $\vari$ and calculate its ground-state eigenvalue and eigenvector.
    \item Add states to $\vari$ via the criterion~(\ref{eq:VHCI_crit}), as follows.  For each state $\staten$, 
    traverse the sorted list of $W_{ijk\ldots}$.
    \begin{enumerate}
        \item If $|W_{i}c_\vn| \ge \eps_1$, add all states $\statem$ that are included in
        $Q_i \staten$.  
        \item If $|W_{ij}c_\vn| \ge \eps_1$, add all states $\statem$ that are included in
        $Q_i Q_j \staten$.  
        \item If $|W_{ijk\ldots}c_\vn| \ge \eps_1$, add all states $\statem$ that are included in
        $Q_i Q_j Q_k \ldots \staten$.  At anharmonic order $p$, there are $O(2^p)$ such states to add.
        \item If $|W_{ijk\ldots}c_\vn| < \eps_1$, break and go on to the next $\staten$.
    \end{enumerate}
    Return to step 3 until the calculation is converged. 
\end{enumerate}

A number of possible convergence criteria can be devised; here, following the
original HCI paper~\cite{Holmes2016}, we consider the calculation converged when the
total number of states added in step 4 is less than 1\% of the current number of
variational states.
Importantly, most elements of $H_{\vm\vn}$ are zero due to the properties of
harmonic oscillators; the states $\statem$ and $\staten$ can only differ by a
few quanta, depending on the maximum anharmonic order of the PES.  The same
would not be true in the basis of product states obtained after a vibrational
self-consistent field procedure~\cite{B.Gerber2002,Chaban1999,Carter1997,Roy2013}, 
which is why we intentionally work
in the noninteracting normal mode basis.

Crucially, the presorting of scaled and/or summed
anharmonic force constants $(W_i, W_{ij}, W_{ijk\ldots})$ combined with the
criterion~(\ref{eq:VHCI_crit}) means that \textit{most states $\statem$ are never even tested for
addition}.  Just like in electronic structure theory, this construction is the
key to the efficiency of VHCI.
Unlike in electronic structure theory, the ground state of the vibrational Schr\"odinger equation, referred to here as the zero-point energy (ZPE), is rarely
of interest by itself.  Following Ref.~\onlinecite{Holmes2017}, we can straightforwardly modify the
above VHCI algorithm to allow the simultaneous calculation of many excited states.  In step 3, we
find all eigenstates of interest (typically the $N_\mathrm{s}$ lowest in energy).  Then we perform the
addition step 4 for each of those states, combining all of their added basis states $\statem$ in the
updated $\vari$.  Clearly, this adds many more basis states at each iteration, but many of them are
duplicates and so the overall variational space is observed to grow sublinearly with $N_\mathrm{s}$.

The above procedure can be applied to a PES expanded to
arbitrary order in the normal mode coordinates.  However, high-order anharmonic interactions
with repeated mode indices will contribute to lower-order excitations beyond the
single and double excitations described in Eqs.~(\ref{eq:single}) and
(\ref{eq:double}) for the case of a quartic PES.  
For example, a sixth-order PES will yield triple excitations due to cubic anharmonicity
\textit{and} fifth-order anharmonicity (when a mode index is repeated), and similarly quadruple
excitations due to quartic and sixth-order anharmonicity.
In this work, we only test one sixth-order PES in Sec.~\ref{ssec:ethylene_oxide};
although we could define new effective force constants for the screening procedure,
e.g.~$W_{ijk}^\prime = W_{ijk} + \sum_k W_{ijkll}$, instead we make the 
approximation $\sum_l W_{ijkll} \approx \max_l W_{ijkll}$.
In other words, we allow fifth-order anharmonicities to propose triple excitations, but only
based on the magnitude of individual anharmonic force constants and not the sum of all such
constants contributing to a given triple excitation.
Our testing suggests that the error incurred is negligible.

\subsection{Epstein-Nesbet perturbation theory}
To the variational energy of each state $E_\mathrm{var}$ we add the
second-order perturbation theory (PT2) correction
\begin{equation}
    \Delta E_2 \approx \sum_\vm^{(\eps_2)} 
    \frac{\left(\sum_\vn 
    H_{\vm\vn}c_\vn\right)^2}{E_\mathrm{var}-H_{\vm\vm}}.
    \label{eq:pt2}
\end{equation}
In exact PT2, the summation over $\vm$ includes all basis states that are
connected to the variational space $\vari$.  For large variational spaces,
this perturbative space is enormous and the cost of the PT2 correction is prohibitive.
To address this, HCI uses the same screening procedure as in the variational stage
to efficiently include only the most important perturbative states~\cite{Holmes2016}.  For VHCI,
we again use criterion~(\ref{eq:VHCI_crit}), with a cutoff $\varepsilon_2 < \varepsilon_1$,
to determine whether basis state $\statem$ should be included in the perturbative space.
When $\varepsilon_2 = 0$, the PT2 calculation is exact within the first-order interacting space.

Throughout this work, we calculate the PT2 correction deterministically as described above,
which ultimately limits the size of the systems that we can accurately study.
In the future we will pursue the stochastic or semistochastic evaluation of Eq.~(\ref{eq:pt2}), as is now
common practice in HCI~\cite{Sharma2017},
in order to study the vibrational structure of even larger systems.

\section{Results}
\label{sec:results}

\begin{table*}[t!]
    \caption{The twenty lowest-energy states from a calculation of the first 70
eigenstates of acetonitrile, with excitation energies given relative to the ZPE.  Mode
assignments are given based on the character of the basis state with the largest
absolute CI coefficient, using mode-numbering from Ref.~\onlinecite{Garnier2016}
and showing multiple assignments when their weights differ by less than 0.1.
VHCI results are shown for two different values of $\eps_1$, which determines
the number of variational basis states, $N_\vari$, and shown with and without
the PT2 correction to the energies (``Var'' and ``Full PT2'', respectively).  We
compare to exact reference values from A-VCI~\cite{Odunlami2017} and to VASCI
with full PT2 correction and $N_\vari$ comparable to our 
$\eps_1=1.0$~\wavenum\ case~\cite{Lesko2019}.  The maximum absolute error (Max.~AE) and 
root mean squared error (RMSE)
relative to A-VCI were calculated across all 70 states.  All energies are
in~\wavenum.
}
    \label{tab:acetylene}
    \centering
    \begin{tabular}{>{\RaggedRight\arraybackslash}p{0.14\textwidth}>{\centering}p{0.12\textwidth}>{\centering}p{0.12\textwidth}>{\centering}p{0.12\textwidth}>{\centering}p{0.12\textwidth}>{\centering\arraybackslash}p{0.12\textwidth}>{\centering\arraybackslash}p{0.12\textwidth}}
\hline\hline
 Method: & \multicolumn{2}{c}{$\eps_1=1.0$} & \multicolumn{2}{c}{$\eps_1=0.1$} & VASCI\cite{Lesko2019} & A-VCI\cite{Odunlami2017} \\
$N_\mathrm{\mathcal{V}}$: & \multicolumn{2}{c}{29 900} & \multicolumn{2}{c}{125 038} & 30 038 & 2 488 511\\\cmidrule[0.4pt](lr){2-3}\cmidrule[0.4pt](lr){4-5}
 & Var & Full PT2 & Var & Full PT2 & Full PT2 & Var \\
\hline
ZPE & 9837.43 & 9837.41 & 9837.41 & 9837.41 & 9837.41 & 9837.41 \\ 
$\omega_{11}$ & 361.01 & 360.99 & 360.99 & 360.99 & 361.01 & 360.99 \\ 
$\omega_{12}$ & 361.01 & 360.99 & 360.99 & 360.99 & 361.01 & 360.99 \\ 
$\omega_{11}+\omega_{12}$ & 723.22 & 723.18 & 723.18 & 723.18 & 723.22 & 723.18 \\ 
$2\omega_{11},~2\omega_{12}$ & 723.25 & 723.19 & 723.18 & 723.18 & 723.23 & 723.18 \\ 
$2\omega_{11},~2\omega_{12}$ & 723.90 & 723.84 & 723.83 & 723.83 & 723.89 & 723.83 \\ 
$\omega_{4}$ & 900.70 & 900.66 & 900.66 & 900.66 & 900.68 & 900.66 \\ 
$\omega_{9}$ & 1034.18 & 1034.13 & 1034.13 & 1034.12 & 1034.14 & 1034.12 \\ 
$\omega_{10}$ & 1034.19 & 1034.13 & 1034.13 & 1034.12 & 1034.14 & 1034.13 \\ 
$\omega_{11}+2\omega_{12}$ & 1086.65 & 1086.56 & 1086.56 & 1086.55 & 1086.64 & 1086.55 \\ 
$2\omega_{1}+\omega_{12}$ & 1086.66 & 1086.56 & 1086.56 & 1086.55 & 1086.64 & 1086.55 \\ 
$3\omega_{11}$ & 1087.88 & 1087.79 & 1087.78 & 1087.78 & 1087.88 & 1087.77 \\ 
$3\omega_{12}$ & 1087.88 & 1087.79 & 1087.78 & 1087.78 & 1087.88 & 1087.77 \\ 
$\omega_{4}+\omega_{11}$ & 1259.89 & 1259.82 & 1259.81 & 1259.81 & 1259.86 & 1259.81 \\ 
$\omega_{4}+\omega_{12}$ & 1259.94 & 1259.83 & 1259.82 & 1259.81 & 1259.86 & 1259.81 \\ 
$\omega_{3}$ & 1389.10 & 1388.99 & 1388.98 & 1388.97 & 1388.99 & 1388.97 \\ 
$\omega_{11}+\omega_{10},~\omega_{9}+\omega_{12}$ & 1394.86 & 1394.71 & 1394.69 & 1394.68 & 1394.76 & 1394.68 \\ 
$\omega_{9}+\omega_{11},~\omega_{10}+\omega_{12}$ & 1394.89 & 1394.71 & 1394.70 & 1394.68 & 1394.79 & 1394.68 \\ 
$\omega_{11}+\omega_{10},~\omega_{9}+\omega_{12}$ & 1395.08 & 1394.93 & 1394.91 & 1394.90 & 1395.00 & 1394.90 \\ 
$\omega_{3}$ & 1397.83 & 1397.70 & 1397.69 & 1397.68 & 1397.77 & 1397.68 \\ 
\hline
Max. AE(70): & 0.61 & 0.14 & 0.05 & 0.01 & 0.50 & -- \\
RMSE(70): & 0.34 & 0.05 & 0.03 & 0.00 & 0.20 & -- \\
\hline\hline
\end{tabular}
\end{table*}

\subsection{Software and simulation details}

All simulations besides those on naphthalene were performed on a 4-core (8-thread) Intel
Core i7-6700 3.4~GHz desktop CPU using up to 16~GB of RAM.  Naphthalene
calculations were performed on a cluster with up to two~12-core Intel Xeon
Gold 6126 2.6~GHz CPUs and using up to 768~GB of RAM.

Our code is based on the Ladder Operator Vibrational Configuration Interaction
package~\cite{lovci} with extensive modifications.
Our VHCI code is available on GitHub~\cite{vhci}.
The Hamiltonian matrix is stored in a sparse format
using the Eigen linear algebra library~\cite{eigen}.
The eigenvalues and eigenvectors of interest are calculated
with the Lanczos algorithm as implemented in the SPECTRA linear algebra 
library~\cite{spectra}.
We verified our code by comparing to the literature and to results
obtained with the PyVCI software package~\cite{Sibaev2016}.

All VHCI calculations were initialized with a basis of all states containing up to two vibrational
quanta in order to ensure that two-quantum overtones and combination
bands, which account for many of the low-lying target states, are present from
the first iteration.  
Our preliminary testing indicates that using a larger initial basis does not qualitatively
improve the convergence of the results.

\subsection{Acetonitrile: A standard benchmark}
\label{ssec:acetonitrile}

We first present results for acetonitrile, CH$_3$CN, a 6-atom,
12-dimensional system that has become one of the canonical benchmarks for new methods
in vibrational structure theory~\cite{Leclerc2014,Brown2016,Baiardi2017,Thomas2018,Lesko2019}.  
We use the quartic PES described in Refs.~\onlinecite{Begue2005,Leclerc2014}, 
for which normal mode frequencies were calculated using CCSD(T)/cc-pVTZ and
higher-order force constants were calculated using B3LYP/cc-pVTZ.  
This PES was also used in highly accurate reference
calculations~\cite{Avila2011,Garnier2016}.
All PESs used in this study are available as
input files on our GitHub repository~\cite{vhci}.

We calculated the energies of the first 70 states of acetonitrile, of which the
first 20 are reported in Table~\ref{tab:acetylene}.  Data for all 70 states can be found in
the supplementary material.  We compare our results to
those obtained using A-VCI~\cite{Garnier2016}, which we consider to
be numerically exact (similar to those obtained with the
nonproduct quadrature approach~\cite{Avila2011}); for reference, the A-VCI
results are obtained with approximately 2.5~million variational states.
Mode assignments here and throughout indicate the character of the product
state with the largest weight in the VCI eigenvector;
for acetonitrile, we use the mode-numbering convention of Ref.~\onlinecite{Garnier2016}.  
To quantitatively assess the accuracy of our results, we report the maximum
absolute error and the root mean squared error of the lowest 70 states. 
In addition to the numerically exact A-VCI results, we also compare to
results obtained recently using VASCI~\cite{Lesko2019}, which is a
selected CI technique that is similar in spirit to VHCI.

We report results for variational VHCI (``Var''), as well as VHCI+PT2 without
VHCI screening of the perturbative space (``Full PT2"); these results are given
for two values of the variational energy cutoff $\eps_1$ that controls the
number of variational states $N_\vari$.  Using $\eps_1=1$~\wavenum\ results in a
variational space with $N_\vari=29\,900$ basis states, which enables
comparison to the largest reported VASCI calculation, with $N_\vari=30\,038$.  
We find that VHCI+PT2 achieves a maximum absolute
error and a root mean squared error that is less than half that of VASCI,
which is somewhat surprising because the CIPSI-style selection criterion
used in VASCI is typically thought to more rigorously identify important states.
However, the precise variational space generated by both VHCI and VASCI can be
tuned by their parameters ($\varepsilon_1$ and the number of core/target states, respectively),
so a direct comparison is not straightforward.
In any event, the accuracy of VHCI is remarkable given the modest
computational cost;  for example, the $\eps_1=1$~\wavenum\ calculation reported here took
less than 3 minutes for the variational stage and
less than 30 minutes for the full PT2 correction (on an 8-core desktop CPU).

\begin{table*}[t!]
    \caption{The sixteen lowest-energy states from a calculation of the first 100
eigenstates of ethylene, with excitation energies given relative to the ZPE.  Calculations were performed using fourth- and sixth-order
truncations of the PES normal mode expansion.  Both VHCI+PT2 calculations were
converged with respect to variational and perturbative basis size.  We compare
to vDMRG\cite{Baiardi2017} for both truncations and to VCI\cite{Sibaev2016}
with up to 8 quanta per product state for the sixth-order truncation.  We also
compare to a calculation from Ref.~\onlinecite{Delahaye2014} which uses an
untruncated version of the PES solved using VCI with a pruned basis
set containing up to 13 quanta per product state.  Assignments are given using the 
mode-numbering convention of Ref.~\onlinecite{Delahaye2014}. All energies are in~\wavenum.}
    \centering
    \begin{tabular}{p{0.11\textwidth}>{\centering}p{0.1\textwidth}>{\centering}p{0.1\textwidth}>{\centering}p{0.1\textwidth}>{\centering\arraybackslash}p{0.1\textwidth}>{\centering\arraybackslash}p{0.1\textwidth}>{\centering\arraybackslash}p{0.1\textwidth}>{\centering\arraybackslash}p{0.1\textwidth}}
\hline\hline
 PES: & \multicolumn{2}{c}{Fourth-order} & 
 \multicolumn{3}{c}{Sixth-order} & Untruncated 
 \\\cmidrule[0.4pt](lr){2-3}\cmidrule[0.4pt](lr){4-6}
 Method: & VHCI+Full PT2 & vDMRG\cite{Baiardi2017} & VHCI+PT2 ($\eps_2=0.01$) & vDMRG\cite{Baiardi2017} & VCI (PyVCI)\cite{Sibaev2015} & Pruned VCI\cite{Delahaye2014} \\\cmidrule[0.4pt](lr){2-2}\cmidrule[0.4pt](lr){4-4}
 & $\eps_1=0.75$ & & $\eps_1=0.5$ & & $N_\mathrm{tot}=8$ & $N_\mathrm{tot}=13$ \\
\hline
 ZPE & 11006.11 & 11006.19 & 11011.61 & 11016.15 & 11011.63 & 	11014.91	\\
 $\omega_{10}$ & 808.61 & 809.03 & 819.99 & 831.17 & 820.11 & 	822.42	\\
 $\omega_{8}$ & 914.87 & 915.29 & 926.33 & 933.47 & 926.45 & 	934.29	\\
 $\omega_{7}$ & 927.87 & 928.31 & 941.65 & 948.26 & 941.78 & 	949.51	\\
 $\omega_{4}$ & 1006.74 & 1007.13 & 1017.45 & 1018.26 & 1017.56 & 	1024.94	\\
 $\omega_{6}$ & 1216.94 & 1217.17 & 1222.16 & 1227.05 & 1222.23 & 	1224.96	\\
 $\omega_{3}$ & 1338.46 & 1338.87 & 1341.95 & 1343.46 & 1342.01 & 	1342.96	\\
 $\omega_{12}$ & 1429.93 & 1430.47 & 1438.31 & 1441.52 & 1438.39 & 	1441.11	\\
 $\omega_{2}$ & 1606.41 & 1622.11 & 1622.78 & 1629.04 & -- & 	1624.43	\\
 $2\omega_{10}$ & 1631.47 & 1625.56 & 1655.21 & 1682.18 & -- & 	1658.39	\\
 $\omega_{8}+\omega_{10}$ & 1718.27 & 1722.77 & 1748.05 & 1770.02 & -- & 	1757.70	\\
 $\omega_{7}+\omega_{10}$ & 1733.98 & 1729.53 & 1766.19 & 1786.99 & -- & 	1778.34	\\
 $\omega_{4}+\omega_{10}$ & 1809.39 & 1810.47 & 1837.68 & 1852.60 & -- & 	1848.61	\\
 $2\omega_{8}$ & 1821.71 & 1826.01 & 1858.36 & 1878.42 & -- & 	1873.73	\\
 $\omega_{7}+\omega_{8}$ & 1821.96 & 1827.96 & 1871.42 & 1886.16 & -- & 	1885.12	\\
 $2\omega_{7}$ & 1850.39 & 1852.23 & 1887.03 & 1906.02 & -- &	1901.61	\\
\hline\hline
\end{tabular}

    \label{tab:ethylene}
\end{table*}

In Table~\ref{tab:acetylene}, we also present results of a larger calculation with
$\eps_1=0.1$~\wavenum, resulting in $N_\vari=125~038$.  We obtain an extremely accurate
spectrum even without the PT2 correction, with a maximum absolute error well
below 0.1~\wavenum.  This variational calculation took less
than 30 minutes on the same 8-core desktop CPU.  This larger
calculation still benefits from minor corrections with PT2, which
takes about three hours and produces a spectrum in which no value differs from the
exact result by more than 0.01~\wavenum\ across all 70 states.  In summary, we
have shown that VHCI can produce near-exact results for a 12-dimensional
system with an extremely small computational effort.

\subsection{Ethylene: Sixth-order potential}
\label{ssec:ethylene}

Next, we study ethylene, C$_2$H$_4$, which is the same size as acetonitrile
(6 atoms, 12 dimensions), but here we use a more realistic \textit{ab
initio} potential energy surface and consider anharmonic expansions up to 
sixth order.  Specifically, we use the PES of
Ref.~\onlinecite{Delahaye2014}, which
was calculated entirely using CCSD(T)/cc-pVQZ,
and we use the PyPES software package~\cite{Sibaev2015} to convert from internal to 
cartesian normal mode coordinates and generate an anharmonic 
expansion up to sixth order.  In contrast to the quartic PES of
acetonitrile that contains 299 nonzero anharmonic force constants, this sixth-order
PES for ethylene contains 2651 force constants, 2375 of which are fifth or
sixth order.  
Here, we compare results and performance when the potential is truncated at 
fourth order and at sixth order.
We calculated the first 100 states; results for the first sixteen states
are shown in Table~\ref{tab:ethylene} and those for the
additional 84 states can be found in the supplementary material.  

Through testing, we confirmed that our VHCI+PT2 results
are converged with $\eps_1=0.75$~\wavenum\ for the quartic case (with full PT2) and 
$\eps_1=0.5$~\wavenum\ for the sixth-order case (with $\eps_2=0.01$~\wavenum), 
resulting in $N_\vari=$ 153\,935 and 161\,338 basis states, 
respectively.  In fact, it was not
possible to go to larger variational spaces for the quartic case because of unphysical 
divergences in the truncated PES. 
Similar behavior of truncated PESs studied at high excitation levels has been observed
before~\cite{Delahaye2014,Sibaev2016,Sibaev2016a}.  

\begin{figure}
    \centering
    \includegraphics{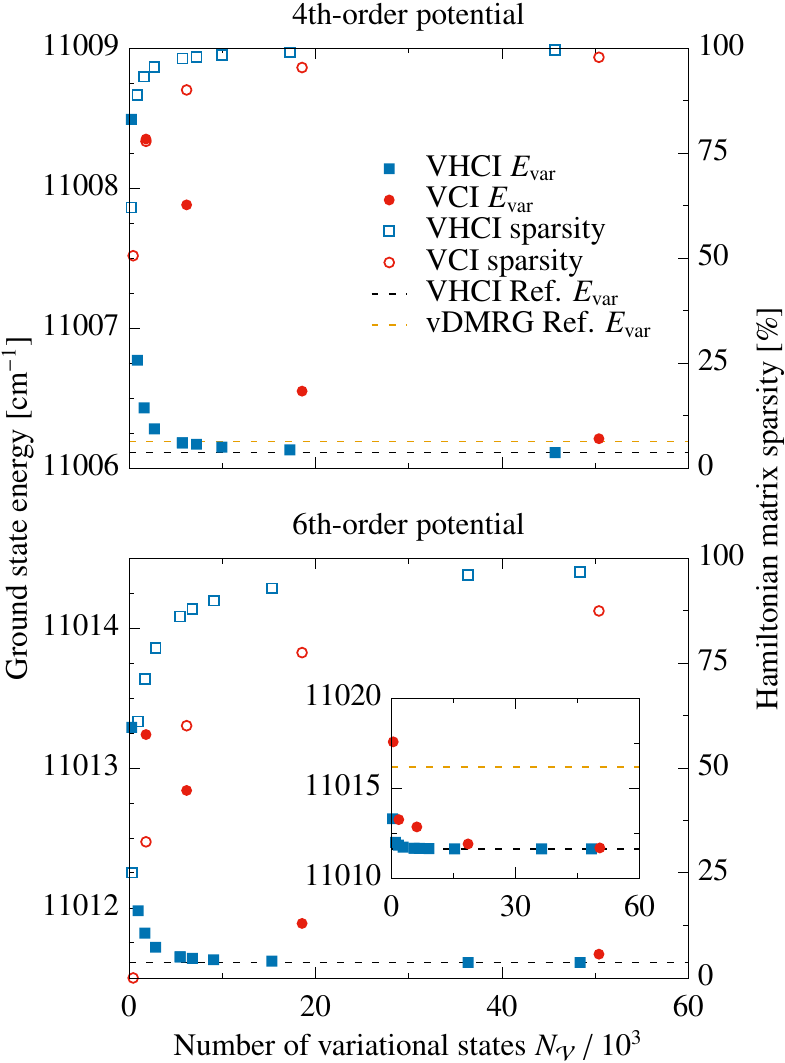}
    \caption{The variational ground-state energy of ethylene as a function of
number of basis states $N_\vari$. We compare the ground state energies of
conventional VCI (solid red circles) and VHCI (solid blue squares) using a
fourth-order (top) and sixth-order (bottom) truncation of the potential.  VHCI
calculations were performed by optimizing for the ground state
($N_\mathrm{s}=1$) and the VCI space was truncated by limiting the
total number of quanta per product state.  The blue dashed
lines represent the converged VHCI ground-state energy, which we consider to be
exact.  The orange dashed lines are the vDMRG values from Ref.~\onlinecite{Baiardi2017}. 
We also present the sparsity of the Hamiltonian matrix (right axis) as a
function of $N_\vari$ for conventional VCI (open red circles) and VHCI (open
blue squares).}
    \label{fig:ethylene}
\end{figure}

In Table~\ref{tab:ethylene}, we also compare our results to those obtained
with the vibrational density-matrix renormalization group (vDMRG)~\cite{Baiardi2017}
using the same fourth- and sixth-order truncated PES.
As discussed in the Introduction, DMRG and selected CI are similarly competitive methods
in electronic structure theory.
Surprisingly, we find that for both truncations, our
energies are noticeably lower than the vDMRG results.  
For the quartic potential, our zero-point energy is 0.8~\wavenum\ lower
than the vDMRG result, while for the sixth-order potential it is 4.5~\wavenum\ lower.  
Figure~\ref{fig:ethylene} shows the convergence of the ZPE as a function of the size of
the variational space, $N_\vari$.  
We see that VHCI (obtained with $N_\mathrm{s}=1$) converges smoothly and quickly;
for the fourth-order potential it exceeds the accuracy of vDMRG with 
about 5000~basis states.  
We also show results obtained with our own VCI code, which
includes basis states according to their total number of vibrational quanta. 
For both the fourth- and sixth-order potentials, VCI converges to the same ZPE as VHCI,
although it requires significantly more basis states for comparable accuracy.
The discrepancy with vDMRG may come from the latter using insufficient bond dimension,
getting trapped in local minima during sweeps, or simply due to slight differences
in the PES.  As a check on the latter, we have also included in Table~\ref{tab:ethylene}
the VCI excitation energies previously reported by the authors of the PyPES software
package~\cite{Sibaev2016} from which the PES parameters were obtained.  The results
are in excellent agreement with our own VHCI+PT2 results, indicating that we are using
the exact same PES as those authors.

In Fig.~\ref{fig:ethylene}, we also plot the percent
sparsity of the Hamiltonian matrix as function of number of variational 
basis states.  We present the
sparsity for VHCI (optimized for the ground state) and conventional VCI.
For both the fourth-order and sixth-order potentials, VHCI produces a much sparser
Hamiltonian matrix than VCI.
These results indicate that VHCI is extremely effective at
capturing the connectivity between important basis states (as demonstrated by the
accurate ground-state energy) while also ignoring the connectivity to less
important basis states (as demonstrated by the increased sparsity).  

In the final column of Table~\ref{tab:ethylene}, we also show results obtained
in Ref.~\onlinecite{Delahaye2014} using a pruned VCI basis with up to 13 vibrational
quanta for the untruncated PES.
In its current form, VHCI requires the truncated normal mode expansion~(\ref{eq:anharm})
and thus comparison to the results obtained for an untruncated PES is important
for assessing the future potential of VHCI.  As can be seen, the agreement improves
significantly when going from the fourth-order PES to the sixth-order PES.
For low-lying states, the disagreement at fourth order is on the order of 10-20~\wavenum\
and at sixth order is on the order of 5~\wavenum, indicating
that the normal mode expansion is sensible and systematically improvable.

\subsection{Ethylene oxide: Convergence and extrapolation}
\label{ssec:ethylene_oxide}

Next, we study ethylene oxide, C$_2$H$_4$O, a molecule with 7
atoms and 15 normal mode degrees of freedom.  Compared to the two previous molecules,
the three additional degrees of freedom make numerically exact VCI calculations much more difficult. 
We use the quartic PES of
B\'egu\'e et al.\cite{Begue2007}, with normal mode frequencies calculated at the
CCSD(T)/cc-pVTZ level and anharmonic force constants calculated using
B3LYP/6-31+G(d,p).  
We use the same version of the PES as 
Refs.~\onlinecite{Odunlami2017,Brown2016,Thomas2018,Lesko2019}, which is available
along with our source code on our GitHub repository~\cite{vhci}.
Like
acetonitrile, ethylene oxide represents a well-studied system that is suitable
for benchmarking new approaches to solving the vibrational structure problem.
In Table~\ref{tab:ethylene_oxide}, we present results for the ten lowest and ten highest
states from a VHCI calculation targeting $N_\mathrm{s}=200$ states.  We performed the calculations with two
different variational cutoff values $\eps_1=2$~\wavenum\ and 1~\wavenum, producing variational
spaces with 132\,163 and 259\,070 basis states, respectively.  We present results
with just the variational stage (``Var") and with heat-bath screened PT2
correction with $\eps_2=0.01$~\wavenum.  

\begin{table*}[ht!]
    \caption{Ten lowest- and ten highest-energy states from the first 200 eigenstates
of ethylene oxide, with excitation energies given relative to the ZPE.  VHCI
results are shown for two different values of $\eps_1$ with no PT2 correction (Var) and with
PT2 at $\eps_2=0.01$~\wavenum.  We also show the first 10
extrapolated energies obtained by a linear fit of $E_\mathrm{tot}$ vs.~$\Delta
E_2$, as shown in Fig.~\ref{fig:extrap}.  We compare all results to numerically exact reference
values from A-VCI~\cite{Odunlami2017} and to VASCI
with full PT2 correction and $N_\vari$ roughly comparable to our 
$\eps_1=2$~\wavenum\ case~\cite{Lesko2019}.  We report the maximum absolute error (Max.~AE) and root mean squared error (RMSE) of the
first 50 states for all methods.  We show the same error metrics for all 200
states for VHCI with $\eps_1=1$~\wavenum\, which is the only case for which all
mode assignments match the exact result.  Assignments use the mode-numbering convention of Ref. \onlinecite{Odunlami2017}. All energies are in~\wavenum.}
    \label{tab:ethylene_oxide}
    \centering
    \begin{tabular}{p{0.11\textwidth}>{\centering}p{0.1\textwidth}>{\centering}p{0.1\textwidth}>{\centering}p{0.1\textwidth}>{\centering}p{0.1\textwidth}>{\centering}p{0.1\textwidth}>{\centering\arraybackslash}p{0.1\textwidth}>{\centering\arraybackslash}p{0.1\textwidth}}
\hline\hline
 Method: & \multicolumn{2}{c}{$\eps_1=2.0$} & \multicolumn{2}{c}{$\eps_1=1.0$} & Linear $\Delta E_2$ & VASCI\cite{Lesko2019} & A-VCI\cite{Odunlami2017} \\
$N_\mathrm{\mathcal{V}}$: & \multicolumn{2}{c}{132 163} & \multicolumn{2}{c}{259 070} & Extrap. & 150 000 & 7 118 214 \\\cmidrule[0.4pt](lr){2-3}\cmidrule[0.4pt](lr){4-5}
 & Var & $\eps_2=0.01$ & Var & $\eps_2=0.01$ & $\eps_2=0.01$ & Full PT2 & Var \\
\hline
ZPE & 12461.63 & 12461.50 & 12461.55 & 12461.48 & 12461.47 & 12461.6 & 12461.47 \\ 
$\omega_1$ & 793.11 & 792.73 & 792.88 & 792.69 & 792.65 & 792.8 & 792.63 \\ 
$\omega_2$ & 822.30 & 821.98 & 822.07 & 821.94 & 821.91 & 822.2 & 821.91 \\ 
$\omega_3$ & 878.62 & 878.33 & 878.41 & 878.30 & 878.28 & 878.4 & 878.27 \\ 
$\omega_4$ & 1017.70 & 1017.24 & 1017.42 & 1017.20 & 1017.15 & 1017.4 & 1017.14 \\ 
$\omega_6$ & 1121.87 & 1121.30 & 1121.53 & 1121.24 & 1121.19 & 1120.8 & 1121.17 \\ 
$\omega_5$ & 1124.37 & 1123.75 & 1124.00 & 1123.70 & 1123.65 & 1123.8 & 1123.62 \\ 
$\omega_7$ & 1146.42 & 1145.84 & 1146.08 & 1145.78 & 1145.73 & 1146.0 & 1145.72 \\ 
$\omega_8$ & 1148.68 & 1148.07 & 1148.31 & 1148.02 & 1147.97 & 1148.3 & 1147.96 \\ 
$\omega_9$ & 1271.43 & 1270.89 & 1271.06 & 1270.83 & 1270.78 & 1271.3 & 1270.78 \\ 
$\omega_1+\omega_5+\omega_9$ & 3175.50 & 3170.13 & 3172.33 & 3169.19 & -- & 3170.9 & 3167.97 \\ 
$\omega_1+\omega_6+\omega_9$ & 3178.91 & 3173.60 & 3175.76 & 3172.68 & -- & 3173.9 & 3171.53 \\ 
$2\omega_4+\omega_8$ & 3181.83 & 3177.51 & 3179.28 & 3176.82 & -- & 3178.0 & 3175.93 \\ 
$\omega_2+\omega_3+\omega_{11}$ & 3187.19 & 3182.12 & 3183.76 & 3181.06 & -- & 3182.1 & 3180.16 \\ 
$3\omega_1+\omega_2$ & 3197.27 & 3191.75 & 3192.18 & 3187.44 & -- & 3190.7 & 3184.85 \\ 
$\omega_1+\omega_8+\omega_9$ & 3198.00 & 3189.61 & 3193.89 & 3190.78 & -- & 3192.7 & 3189.65 \\ 
$\omega_2+\omega_5+\omega_9$ & 3205.08 & 3200.30 & 3202.16 & 3199.48 & -- & -- & 3198.56 \\ 
$\omega_2+\omega_6+\omega_9$ & 3206.11 & 3201.03 & 3202.96 & 3200.15 & -- & 3200.9 & 3199.21 \\ 
$\omega_1+\omega_7+\omega_9$ & 3210.44 & 3205.96 & 3207.74 & 3205.19 & -- & 3206.7 & 3204.35 \\ 
$2\omega_1+2\omega_2$ & -- & -- & 3218.48 & 3214.59 & -- & -- & 3212.77 \\
\hline
Max AE(50): & 4.00 & 0.98 & 2.09 & 0.50 & 0.06 & 2.2 & -- \\
RMSE(50): & 2.40 & 0.51 & 1.22 & 0.26 & 0.03 & 0.8 & -- \\
\hline
Max. AE(200): & -- & -- & 7.33 & 3.23 & -- & -- & -- \\
RMSE(200): & -- & -- & 2.96 & 0.84 & -- & -- & -- \\
\hline
Core hours: & 6.9 & 20.4 & 28.2 & 59.8 & -- & 67.1 & 1756.4 \\
Cores: & 8 & 8 & 8 & 8 & -- & 2 & 24 \\
\hline\hline
\end{tabular}

\end{table*}

\begin{figure*}[t!]
    \centering
    \includegraphics{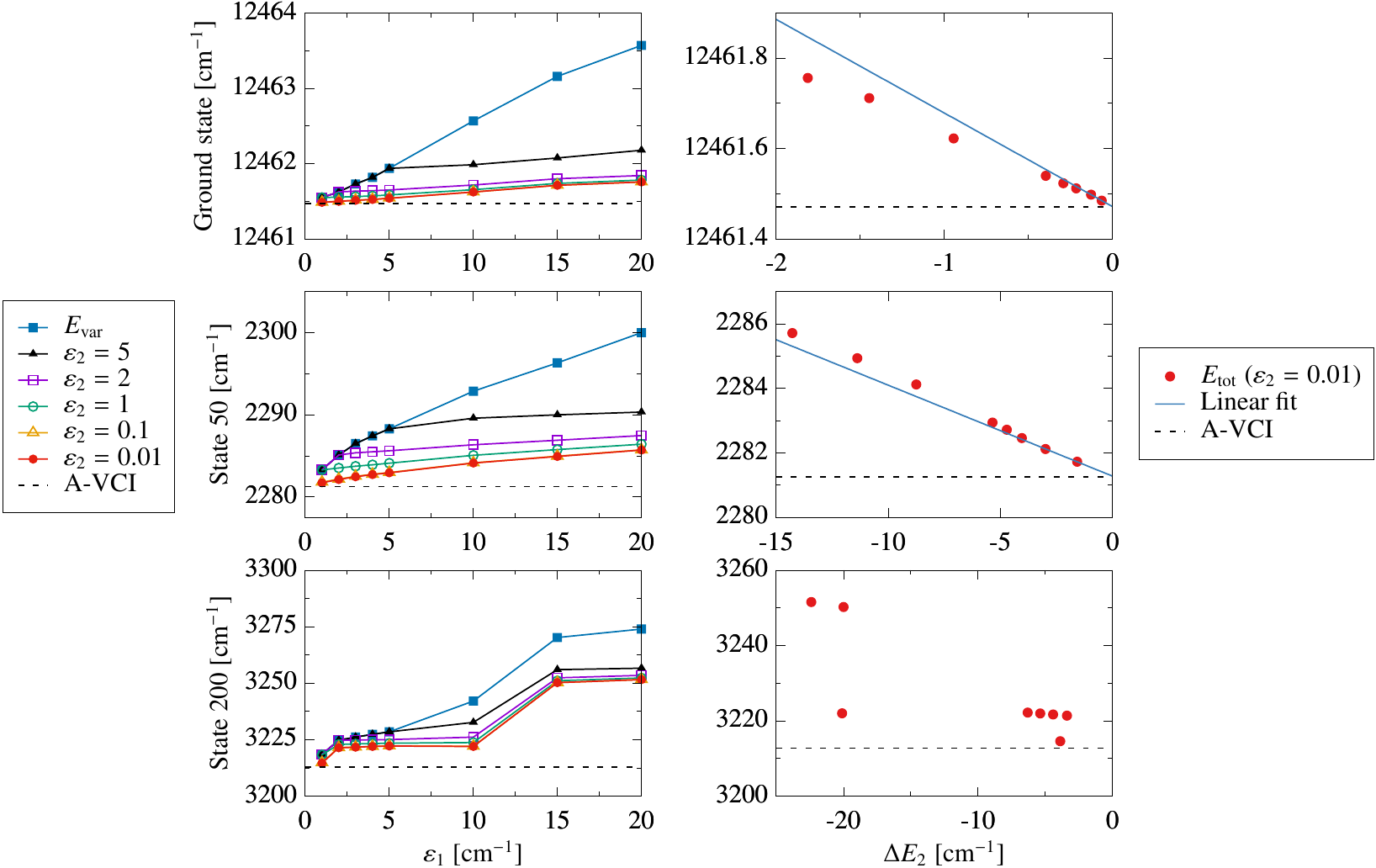}
    \caption{VHCI energy of the ground state (top row), the 50th state
(middle row) and 200th state (bottom row) of ethylene oxide.  The left-hand 
column shows the energy as a function of the variational heat-bath cutoff
parameter $\eps_1$ and includes the variational energy $E_\mathrm{var}$ (blue
squares) and the total energy $E_\mathrm{tot}=E_\mathrm{var}+\Delta E_2$ for
various values of the perturbative heat-bath screening parameter $\eps_2$.  The
black dashed lines are the converged A-VCI values for each state, which can be
considered to be exact.  In the right-hand column, we present the total energy
$E_\mathrm{tot}$ (red circles) with converged PT2 ($\eps_2=0.01$~\wavenum) as a
function of the perturbative correction $\Delta E_2$.  Note that we use a
smaller y-range because the energy varies less with respect to $\eps_1$ when the
PT2 stage is converged.  We also show the linear fits (blue lines) of the most
accurate two points ($\eps_1=1$ and 2~\wavenum) for the ground state and state
50, which were used to find the extrapolated energies presented in Table~\ref{tab:ethylene_oxide}.}
    \label{fig:extrap}
\end{figure*}

On the left-hand side of Fig.~\ref{fig:extrap}
we plot the energy of the first, 50th, and 200th state with respect to the 
variational cutoff $\eps_1$, for a variety of values of the perturbative screening parameter
$\eps_2$.  We see that the curves with $\eps_2=1$~\wavenum, 0.1~\wavenum, and 0.01~\wavenum\ agree 
very
closely for all $\eps_1$, with $\eps_2=0.1$~\wavenum\ and 0.01~\wavenum\ being indistinguishable,
confirming convergence with respect to $\eps_2$.  
In Table~\ref{tab:ethylene_oxide} and Fig.~\ref{fig:extrap}, we compare our results to
A-VCI calculations~\cite{Odunlami2017} that were
obtained from an optimized variational space containing over 7~million basis states, which
we take to be numerically exact.
In Fig.~\ref{fig:extrap},
we see that all variational calculations tend monotonically toward the exact energies as
$\eps_1$ becomes small.  Addition of the PT2 correction leads to more rapid
convergence with respect to $\eps_1$.  For example, for the ground state we see
that results obtained with $\eps_1=20$~\wavenum\ and converged PT2 gives an answer that is closer to exact than
that with $\eps_1=5$~\wavenum\ and no PT2 correction.  The benefits of perturbation theory become
less pronounced in high-lying excited states, due to their large multiconfigurational character.  
In fact, for the 200th state, our calculation only produces the correct band assignment for
$\eps_1\le 1$~\wavenum.  Table~\ref{tab:ethylene_oxide} shows the band assignment
(obtained from the variational calculation) following
the mode labeling convention of Ref.~\onlinecite{Odunlami2017} and we leave the
energy blank if we do not have an assignment that corresponds to the exact
result.  The 200th state is omitted for all methods except
VHCI at $\eps_1=1$~\wavenum\ and the A-VCI reference; several of the other ten highest
excitations are also omitted from the VASCI results~\cite{Lesko2019}.  
Because the incorrect band assignments of high-lying states
prevents us from comparing level-by-level with exact results, we include
statistics on just the first 50 states.  Variational VHCI produces good agreement for the first 50
states, with a RMSE on the order of 1-2~\wavenum~and maximum AE below 5~\wavenum;
the addition of PT2 yields accuracy better than 1~\wavenum\ for both $\eps_1=1$~\wavenum\ 
and 2~\wavenum.  In comparison,
VASCI+Full PT2 produces a maximum AE of 2.2~\wavenum\ and RMSE of 0.8~\wavenum\ for the first 50
levels compared to the exact results.  All of the correct band assignments are
present in the first 50 states of both VHCI and VASCI at $N_\vari=150\,000$, but
the closely-spaced states 47 and 48 have the wrong energetic ordering for VASCI
and for VHCI with $\eps_1=2$~\wavenum.
For
$\eps_1=1$~\wavenum, we obtain all 200 states with the correct assignments, enabling
directe comparison to A-VCI.  We see very good agreement over all
states and VHCI+PT2 has a maximum AE around 3~\wavenum\ and RMSE below 1~\wavenum.  
Remarkably, this calculation took less than 8 hours on an 8-core desktop CPU.
As a rough comparison to other methods, we also present the timings for all calculations 
in Table~\ref{tab:ethylene_oxide}.

Finally, following standard HCI practice in electronic structure, we attempt to
approximate the exact energies via extrapolation.  
On the right-hand side of Fig.~\ref{fig:extrap}, we plot the VHCI+PT2 energy
of each state as a function of the PT2 correction $\Delta E_2$ seeking to
extrapolate to $\Delta E_2=0$, following Ref.~\onlinecite{Holmes2017}.
We see that extrapolation is reasonable and successful for low-lying states,
but not for high-lying states.
In general, extrapolation of high-lying states is difficult because
the level spacing becomes very small and the states are highly multiconfigurational,
such that tracking a single level as $\eps_1$ changes is challenging.  
We performed extrapolation for the first 50 states using a linear fit to results
obtained with $\eps_1=1$~\wavenum\ and 2~\wavenum, as shown graphically in the
right-hand sides of Fig.~\ref{fig:extrap} for the first and 50th states;
we tested other polynomial fits and found linear extrapolation to be the
simplest and best perfoming, although other schemes could be considered
in the future.
In Table~\ref{tab:ethylene}, we present the extrapolated energies of the
first ten states as well as the statistics of the first 50 states; all energies can
be found in the supplementary material.
The results are nearly exact, with a maximum AE below 0.1~\wavenum\ for all 50
states, and obviously require no additional computing effort. 
We conclude that linear extrapolation of high-quality VHCI+PT2 energies is a
powerful way to achieve nearly exact energies for the ground state and many 
low-lying excited states.

\begin{table*}[t!]
    \caption{Ten lowest- and ten highest-energy states from a calculation of
the first 128 eigenstates of naphthalene, with excitation energies given relative to the
ZPE.  VHCI results are shown for two different values of $\eps_1$ with no PT2 correction
(Var) and with PT2 at $\eps_2=0.2$~\wavenum.  We compare to reference values
from the smallest and largest HI-RRBPM calculations in Ref.~\onlinecite{Thomas2018} 
(``A" and ``G" respectively) as well as to
VASCI~\cite{Lesko2019} with no PT2 correction and $N_\vari$ similar to our
$\eps_1=1.5$~\wavenum\ case.  Assignments are given using the mode-numbering
convention of Ref.~\onlinecite{Cane2007}. All energies are in~\wavenum}
    \label{tab:naphthalene}
    \centering
    \begin{tabular}{p{0.11\textwidth}>{\centering}p{0.1\textwidth}>{\centering}p{0.1\textwidth}>{\centering}p{0.1\textwidth}>{\centering}p{0.1\textwidth}>{\centering}p{0.1\textwidth}>{\centering}p{0.1\textwidth}>{\centering\arraybackslash}p{0.1\textwidth}>{\centering\arraybackslash}p{0.1\textwidth}}
\hline\hline
Method: & \multicolumn{2}{c}{$\eps_1=1.5$} & \multicolumn{2}{c}{$\eps_1=1.0$} & \multicolumn{2}{c}{VASCI 
\cite{Lesko2019}
} &  \multicolumn{2}{c}{HI-RRBPM 
\cite{Thomas2018}
}  \\\cmidrule[0.4pt](lr){6-7}\cmidrule[0.4pt](lr){8-9}
$N_\mathrm{\mathcal{V}}$: & \multicolumn{2}{c}{1\,322\,334} & \multicolumn{2}{c}{2\,270\,672} &  1\,000\,000 & 1\,500\,000 & A & G \\\cmidrule[0.4pt](lr){2-3}\cmidrule[0.4pt](lr){4-5}
 & Var & $\eps_2=0.2$ & Var & $\eps_2=0.2$ & Var & Var & Var & Var \\
\hline
ZPE & 31772.71 & 31764.77 & 31769.90 & 31764.34 & 31774.4 & 31773.6 & 31782.20 & 31766.03 \\ 
$\omega_{48}$ & 168.17 & 165.20 & 166.89 & 164.63 & 166.4 & 166.3 & 165.84 & 164.60 \\ 
$\omega_{13}$ & 182.57 & 179.28 & 181.28 & 178.79 & 179.9 & 179.9 & 184.90 & 178.18 \\ 
$2\omega_{48}$ & 335.22 & 329.48 & 333.17 & 328.69 & 336.4 & 336.1 & 338.21 & 329.41 \\ 
$\omega_{13}+\omega_{48}$ & 349.25 & 342.84 & 346.48 & 341.76 & 349.5 & 349.4 & 365.68 & 342.02 \\ 
$\omega_{24}$ & 358.59 & 356.26 & 357.83 & 355.97 & 361.4 & 363.8 & 372.86 & 354.44 \\ 
$2\omega_{13}$ & 362.91 & 357.23 & 360.75 & 356.49 & 363.4 & 366.6 & 397.32 & 357.66 \\ 
$\omega_{16}$ & 392.38 & 389.05 & 391.19 & 388.64 & 394.1 & 396.3 & 405.35 & 387.71 \\ 
$\omega_{28}$ & 468.62 & 464.50 & 467.08 & 463.93 & 470.9 & 475.3 & 468.20 & 463.47 \\ 
$\omega_{47}$ & 477.41 & 472.97 & 475.83 & 472.45 & 479.7 & 483.1 & 477.10 & 472.41 \\ 
$3\omega_{48}$ & 503.57 & 493.98 & 499.45 & 492.38 & 502.5 & 501.8 & 506.60 & 495.50 \\ 
$\omega_{13}+\omega_{23}$ & 978.99 & 972.19 & 976.88 & 971.60 & 991.6 & 1007.1 & 1091.20 & 974.99 \\ 
$\omega_{24}+\omega_{36}$ & 982.41 & 977.49 & 981.37 & 977.64 & 1006.6 & 1015.4 & 1099.87 & 982.87 \\ 
$\omega_{12}+\omega_{24}$ & 984.01 & 978.50 & 982.39 & 977.99 & 1002.2 & 1015.7 & 1100.56 & 985.06 \\ 
$\omega_{9}+\omega_{28}$ & 984.89 & 978.67 & 982.83 & 978.03 & 997.7 & 1010.7 & 1098.74 & 981.31 \\ 
$\omega_{44}+\omega_{47}$ & 986.66 & 979.81 & 984.23 & 979.12 & 1001.8 & 1013.3 & 1106.07 & 987.92 \\ 
$\omega_{9}+\omega_{47}$ & 993.93 & 987.08 & 991.49 & 986.44 & -- & -- & 1121.57 & 994.66 \\ 
$\omega_{35}$ & 1013.18 & 1008.63 & 1011.79 & 1008.21 & -- & -- & 1134.08 & 1011.99 \\ 
$\omega_{16}+\omega_{36}$ & 1017.62 & 1011.96 & 1015.91 & 1011.51 & -- & -- & 1138.09 & 1013.55 \\ 
$2\omega_{44}$ & 1017.51 & 1012.86 & 1016.19 & 1012.70 & -- & -- & 1138.46 & 1015.09 \\ 
$\omega_{9}+\omega_{44}$ & 1024.79 & 1020.17 & 1023.43 & 1019.82 & -- & -- & 1148.79 & 1018.96 \\
\hline
Max AE(25): & 11.78 & 2.85 & 6.32 & 3.69 & 16.1 & -- & 54.64 & -- \\
RMSE(25): & 6.62 & 1.35 & 4.18 & 1.68 & 9.0 & -- & 30.44 & -- \\
\hline
Core hours: & 1234.4 & 2004.0 & 2620.8 & 3993.4 & 1584.7 & 1834.9 & 1167.4 & 63590.4 \\
Cores: & 20 & 20 & 20 & 20 & 40 & 40 & 128 & 64 \\
\hline\hline
\end{tabular}

\end{table*}

\subsection{Naphthalene: A 48-dimensional system}
\label{ssec:naphthalene}
As a final test of VHCI, we consider naphthalene, 
C$_{10}$H$_8$, with 18 atoms and 48 normal modes, making it more than three times
larger than any of the previous test systems. 
We use the quartic PES of Can\'e et al.~\cite{Cane2007}, which was calculated at the 
B97-1/TZ2P level of theory, and includes
4125 nonzero anharmonic force constants.  Large variational calculations were previously 
performed with this PES using the Hierarchical Intertwined
Reduced-Rank Block Power Method (HI-RRBPM)~\cite{Thomas2018}.  
We compare to the affordable HI-RRBPM~(A) results and the most expensive HI-RRBPM~(G)
results, which we consider to be the most accurate.
Finally, we compare to variational VASCI
calculations from Ref.~\onlinecite{Lesko2019} with 1~million and
1.5~million basis states, obtained using 25\,000 and
15\,000 core states, respectively.  Following Refs.~\onlinecite{Thomas2018,Lesko2019}, we
calculate the 128 lowest states of naphthalene, of which the ten lowest and
ten highest are presented in Table~\ref{tab:naphthalene}; results for all states can
be found in the supplementary material.  We show results with
VHCI variational cutoff values $\eps_1=1.5$~\wavenum\ and 1~\wavenum, producing
approximately 1.3~million and 2.3~million basis states, respectively.  Although a
fully converged PT2 correction is intractable, we
calculate an approximate PT2 correction with heat-bath screening;
we used $\eps_2=0.2$~\wavenum, producing a
perturbative space containing approximately 15~million states.

For low-lying states, the agreement between variational VHCI and HI-RRBPM~(G)
is very good.  At comparable computational cost, the accuracy of variational
VASCI and VHCI is similar.  The PT2 correction to VHCI
produces a significant improvement of low-lying energies, which now
match HI-RRBPM~(G) to an accuracy of about 1~\wavenum.

We calculated the maximum absolute error and RMSE with respect to HI-RRBPM for
the first 25 eigenstates.  We matched the states to Ref.~\onlinecite{Thomas2018} and
Ref.~\onlinecite{Lesko2019} according to their mode assignments.  Variational VHCI
achieves marginally closer agreement to the large HI-RRBPM(G) calculation than
VASCI.  Perturbation theory provides a noticeable improvement over the
variational result.  Curiously, VHCI+PT2 produces a more accurate result for the
smaller variational space, indicating that the calculation is probably not converged with
respect to $\eps_2$. Indeed, ideally the PT2 correction should be calculated with $\eps_2 \ll \eps_1$, 
but even $\eps_2=0.2$~\wavenum\ produces an accurate result for
low-lying states.  As a convergence test, we also performed a large VHCI
calculation that optimizes only the ground state (not shown), and obtained a variational ZPE that is at least 4~\wavenum\ lower
than the HI-RRBPM~(G) answer;
therefore, some of our VHCI results that are lower than HI-RRBPM~(G) may actually be more accurate, which would explain some of the
discrepancies seen in the comparison.

For the high-lying states, which are much harder to converge as we discussed above,
we only present results for those states that
match the mode assignment from HI-RRBPM~(G) and we do not attempt to calculate error statistics.
For states with matching assignments,
we see good agreement between variational VHCI and HI-RRBPM~(G),
especially at $\eps_1=1$~\wavenum.  
The PT2 correction is not as helpful for high-lying states as it is for low-lying ones,
because the variational space of the low-lying states is tightly converged
and additional corrections are well-captured by
perturbation theory.  In some cases, the PT2 correction worsens the agreement with HI-RRBPM~(G),
although this may be a result of an incorrect mode assignment inside a dense spectrum
of excited states.
We do not attempt any extrapolation
because $\eps_2=0.2$~\wavenum\ is not small enough to converge the PT2 correction;
a stochastic PT2 implementation~\cite{Sharma2017} would be clearly beneficial.
Overall, we are confident that the eigenvalue spectrum obtained from our
VHCI calculations is accurate enough to be useful in real-world applications.
Comparison of the overall CPU times, presented at the bottom of Table~\ref{tab:naphthalene}, 
underscores the competitiveness of VHCI as a means of
accurately solving the vibrational Schr\"odinger equation on high-dimensional anharmonic
potentials.

\section{Conclusions}
\label{sec:conc}
We have introduced vibrational heat-bath configuration interaction based on the 
original principles of HCI~\cite{Holmes2016}, but with
adaptations for vibrational Hamiltonians.  
VHCI+PT2 performed well on our four test molecules, achieving quantitative accuracy
for the 12-dimensional systems acetonitrile and ethylene, while outperforming VASCI
and vDMRG.  VHCI also performed well on larger systems, especially for low-lying
states.  High-lying states are a challenge due to their highly
multi-configurational character and dense energetic spacing.  Convergence of
high-lying states could be improved by implementing a state-specific algorithm
\cite{Rauhut2007,Neff2009,Carbonniere2010}.  

In future work, the implementation of semistochastic PT2~\cite{Sharma2017,Holmes2017,Li2018} will
be critical for studying even larger systems.
Furthermore, VHCI can be straightforwardly extended to
efficiently calculate spectroscopic intensities based on the dipole moment surface
\cite{Baraille2001,Burcl2003,Carbonniere2010}.  Consideration of spectroscopic
activity can also be used to target eigenstates more efficiently, as recently
implemented in A-VCI \cite{LeBris2020}.
Finally, it will be interesting to apply VHCI to more strongly anharmonic systems,
such as molecular clusters~\cite{Kim1994,Yu2019} or floppy molecules~\cite{Bacic1989,Bramley1994,Roy2013},
where truncated expansions of the PES may not be sufficient.

\section*{Supplementary material}
See the supplementary material for VHCI energies and assignments of all states calculated
for the four molecules considered in this work.
 
\section*{Acknowledgements}
We thank Dr.~James Brown and Dr.~Marc Odunlami for providing us with accurate
expanded potential energy surfaces and reference energies for acetonitrile and
ethylene oxide and Prof.~Sandeep Sharma for helpful discussions.  
J.H.F.~was supported in part by the National Science Foundation Graduate
Research Fellowship under Grant No.~DGE-1644869.  
We acknowledge computing resources from Columbia University’s Shared Research
Computing Facility project, which is supported by NIH Research Facility
Improvement Grant 1G20RR030893-01, and associated funds from the New York State
Empire State Development, Division of Science Technology and Innovation (NYSTAR)
Contract
C090171, both awarded April 15, 2010.
The Flatiron Institute is a division of
the Simons Foundation.

\section*{Data availability statement}
The software and data that support the findings of this study are openly available in
the supplementary material and in our GitHub repository at \url{https://github.com/berkelbach-group/VHCI} with digital object identifier \href{https://doi.org/10.5281/zenodo.4116070}{10.5281/zenodo.4116070}.

\bibliography{VHCI.bib,software.bib}

\end{document}